\documentclass[journal]{IEEEtran}

\usepackage[T1]{fontenc}
\usepackage[latin9]{inputenc}
\usepackage{color}
\usepackage{float}
\usepackage{amsthm}
\usepackage[nosumlimits]{amsmath}
\usepackage{amssymb}
\usepackage{multirow}
\usepackage{overpic}

\makeatletter

\floatstyle{ruled}
\newfloat{algorithm}{tbp}{loa}
\providecommand{\algorithmname}{Algorithm}
\floatname{algorithm}{\protect\algorithmname}

\allowdisplaybreaks

\usepackage{mathtools}

\usepackage{amsfonts}
\usepackage{cite}
\usepackage{array}
\usepackage{algorithm}
\usepackage{algorithmic}
\usepackage{graphicx}

\newcommand*{\rom}[1]{\expandafter\@slowromancap\romannumeral #1@}

\makeatother

\mathchardef\mhyphen="2D

\newcommand{\rev}[1]{{\color{black}#1}}

\usepackage{siunitx} 
\usepackage{subfig}
\usepackage[font = footnotesize]{caption}

\usepackage{url}
\usepackage{hyperref}

\begin{document}

\title{Mobile Communications Beyond \SI{52.6}{GHz}: Waveforms, Numerology, and Phase Noise Challenge}

\author{
\IEEEauthorblockN{Toni Levanen, Oskari Tervo, Kari Pajukoski, Markku Renfors, and Mikko Valkama %
}
\thanks{T. Levanen, M. Renfors, and M. Valkama are with Department of Electrical Engineering, Tampere University, Finland (firstname.lastname@tuni.fi)}
\thanks{O. Tervo and K. Pajukoski are with Nokia Bell Labs, Oulu, Finland (firstname.lastname@nokia-bell-labs.com)}
\thanks{This article contains complementary electronic materials, available at \newline {\color{blue}\texttt{\href{http://www.tut.fi/5G/WCM2019B50GHz/}{http://www.tut.fi/5G/WCM2019B50GHz/}}}}
\vspace{-2mm}
}

\maketitle

\begin{abstract}

In this article, the first considerations for the 5G New Radio (NR) physical layer evolution to support beyond \SI{52.6}{GHz} communications are provided. In addition, the performance of both OFDM based and DFT-s-OFDM based networks are evaluated with special emphasis on the phase noise (PN) induced distortion. It is shown that DFT-s-OFDM is more robust against PN under 5G NR Release 15 assumptions, namely regarding the supported phase tracking reference signal (PTRS) designs, since it enables more effective PN mitigation directly in the time domain. To further improve the PN compensation capabilities, the PTRS design for DFT-s-OFDM is revised, while for the OFDM waveform a novel block PTRS structure is introduced, providing similar link performance as DFT-s-OFDM with enhanced PTRS design. We demonstrate that the existing \rev{5G NR Release 15} solutions can be extended to support efficient mobile communications at \SI{60}{GHz} carrier frequency with the enhanced PTRS structures. In addition, DFT-s-OFDM based downlink for user data could be considered for beyond \SI{52.6}{GHz} communications to further improve \rev{system power efficiency and} performance with higher order modulation and coding schemes. \rev{Finally, network link budget and cell size considerations are provided, showing that at certain bands with specific transmit power regulation, the cell size can eventually be downlink limited.}
\end{abstract}
\vspace{-4mm}

\section{Introduction}

The frequencies beyond \SI{52.6}{GHz} contain very large spectrum opportunities and will thus facilitate many high capacity use cases, such as integrated access and backhaul, ultra-high data rate mobile broadband, device-to-device communications, and industrial internet-of-things applications, as envisioned in \cite{3GPPTR38807}. Together, these would enable completely new applications for augmented or virtual reality services, factory automation, and intelligent transport systems. Operation at the millimeter waves (mmWaves) has, however, many differences compared to the lower frequencies\rev{\cite{2019:B:Holma:5GTechnology,2017:J:Shafi:5GOverview}}. Firstly, the severely increased path loss (PL) implies that directional antenna arrays with large number of antenna elements are needed. Beam-based operation with narrow beams results in more complex channel access mechanisms, more exhaustive beam training and refinement procedures, and mainly line-of-sight (LOS) communications. Another major factor is the decreased power efficiency of power amplifiers (PAs). Specifically, PAs operating at mmWaves have typically lower output powers while being also more non-linear compared to the PAs at the traditional below \SI{3}{GHz} bands \cite{2019:J:Shakib:CMOSPAs5G}. This implies that higher power back-off is required which directly decreases \rev{the system power efficiency and coverage}. It is also well known that orthogonal frequency division multiplexing (OFDM) signals have larger peak-to-average-power ratio (PAPR) than discrete Fourier transform (DFT) spread OFDM (DFT-s-OFDM) \rev{\cite{2019:B:Holma:5GTechnology}}, especially at lower modulation orders. \rev{This raises the importance of supporting DFT-s-OFDM in both downlink (DL) and uplink (UL) in the beyond \SI{52.6}{GHz} networks \rev{for enhanced coverage and power efficiency} -- an important issue that is addressed in this article.}

\begin{table*}[]
    \setlength{\tabcolsep}{3pt}
    \renewcommand{\arraystretch}{1.3}
    \caption{Current spectrum availability in various countries between frequencies \SI{52.6}{GHz} and \SI{100}{GHz} \cite{3GPPTR38807}, including indicators for unlicensed (U) or licensed (L) spectrum access and for allowed use cases of mobile (M) or fixed (F) point-to-point communications.}
    \centering
    \begin{tabular}{|l|c|c|c|c|c|c|c|c|c|c|c|c|c|}
        \hline
        \multirow{2}{*}{\textbf{Country/Region}} & \multicolumn{13}{c|}{\textbf{Frequency [GHz]}} \\ \cline{2-14}
        & 52.6-57 & 57-59 & 59-64 & 64-66 & 66-$71^{1)}$ & 71-76 & 76-81 & 81-86 & 86-92 & 92-94 & 94-94.1 & 94.1-95 & 95-100 \\ \hline
        Europe  & & \multicolumn{3}{c|}{U;M} & U/L;M & L;F & & L;F & & \multicolumn{4}{c|}{L;F} \\ \hline
        South Africa & & \multicolumn{3}{c|}{U;M} & U/L;M & L;F & & L;F & & & & &\\ \hline
        USA & & \multicolumn{4}{c|}{U;M} & L;F/M& & L;F/M & & L;F/M & & L;F/M & \\ \hline
        Canada & & \multicolumn{2}{c|}{U;M} & & U/L;M & L;F & & L;F & & & & & \\ \hline
        Brazil & & \multicolumn{2}{c|}{U;M} & & U/L;M & L;F & & L;F & & & & & \\ \hline
        Mexico & & \multicolumn{2}{c|}{U;M} & &  U/L;M& L;F & & L;F & & & & & \\ \hline
        China & & & U;M & & U/L;M & & & & & & & &\\ \hline
        Japan & & \multicolumn{3}{c|}{U;M} & U/L;M & L;F & & L;F & & & & &\\ \hline
        Singapore & & \multicolumn{3}{c|}{U;M} & U/L;M & L;F & & L;F & & & & &\\ \hline
        \multicolumn{14}{|l|}{1) \rev{Allowed for administrations wishing to implement the terrestrial component of IMT, by RESOLUTION COM4/7 (WRC-19).}} \\  \hline
    \end{tabular}
    \label{tab:spectrumAvailability}
\end{table*}

\rev{Another important implementation challenge at mmWaves is the oscillator phase noise (PN).} Already in the 3GPP Release 15 (Rel-15) of \rev{5G} New Radio (NR)  \cite{2017:J:Parkvall:5GNR,2017:J:Shafi:5GOverview,3GPPTS38300}, PN was considered as a part of the air interface design. More specifically, 3GPP defined the so-called phase tracking reference signals (PTRSs), which allow the receiver (RX) to estimate PN from known reference symbols and compensate it before decoding the data. \rev{However, the currently supported designs may not be sufficient to guarantee good performance in beyond \SI{52.6}{GHz} communications -- an aspect that is explicitly addressed in this article.} This is mainly due to the fact that PN increases by \SI{6}{dB} for every doubling of the carrier frequency \rev{\cite{2000:J:Lee:OscillatorTutorial}}. 

\rev{In this paper, after some short considerations on the available spectrum bands for the beyond \SI{52.6}{GHz} networks, the evolution of the physical layer (PHY) numerology in  terms of the subcarrier spacing (SCS), slot duration and potential channel bandwidths is addressed and discussed. Then, the important oscillator phase noise challenge is studied, with specific emphasis on new and novel PTRS structures. This is followed by example radio link performance evaluations at \SI{60}{GHz}, considering both OFDM and DFT-s-OFDM waveforms, where the envisioned new PHY numerologies as well as the different PTRS designs are deployed, while also comparing against the Rel-15 NR specification based network as baseline.It is shown that the DFT-s-OFDM waveform is more robust against PN induced distortion and can operate with smaller SCS than OFDM, especially with higher modulation orders. This is due to the time domain group-wise PTRS design used with DFT-s-OFDM, which allows to track the time varying PN realization within a DFT-s-OFDM symbol. We also show that DFT-s-OFDM based radio link performance can be further improved by designing new PTRS configurations with only modest increase in the PTRS overhead. In addition, to improve the OFDM performance, a novel block PTRS design is adopted, allowing to nearly achieve the link performance of DFT-s-OFDM with enhanced PTRS configuration. Finally, the important cell size and the associated UL/DL link budget aspects are addressed, considering again both OFDM and DFT-s-OFDM waveforms, while also comparing PA technology limited and effective isotropic radiated power (EIRP) limited systems. It is shown that at EIRP limited bands, the network coverage may actually be downlink limited.}

\vspace{-1mm}
\rev{\section{Spectrum Prospects at Beyond \SI{52.6}{GHz} Bands}}
\label{section:newBands}

The current spectrum availability between frequencies \SI{52.6}{GHz} and \SI{100}{GHz} \cite{3GPPTR38807} is illustrated at high level in Table \ref{tab:spectrumAvailability}. Regarding the unlicensed access based mobile communications, global spectrum is available at frequency range \SI{59}{GHz} -- \SI{64}{GHz}. Furthermore, the best availability of unlicensed access for mobile communications is in USA, where the wide frequency range of \SI{57}{GHz} -- \SI{71}{GHz} is providing a total of \SI{14}{GHz} of bandwidth. As there are uncertainties when the other beyond \SI{52.6}{GHz} frequency bands are available for mobile communications, there is a strong motivation to study the extension of the current Rel-15 solutions to operate also in the frequency range of \SI{57}{GHz} -- \SI{71}{GHz}, commonly called the  \SI{60}{GHz} band(s). It should be highlighted, however, that the direct extension of current Rel-15 operation is not a suitable long term solution when aiming to cover a wider range of carrier frequencies beyond \SI{52.6}{GHz}. At the moment, the upper limit of the 3GPP related studies is set to \SI{114.25}{GHz} \cite{3GPPTR38807}, but even higher carrier frequencies could be envisioned in the near future, and thus a study on a new common waveform for beyond \SI{52.6}{GHz} communications is of large interest. \rev{Potential physical layer numerology solutions to this direction are addressed and discussed in Section III.}

In general, when looking at frequencies above \SI{71}{GHz}, we note that they are mainly reserved for fixed, point-to-point communications, except in USA where both mobile and fixed communications are allowed. Considering the regulation in Europe for the frequency range \SI{71}{GHz} -- \SI{100}{GHz}, we can see that up to \SI{18}{GHz} aggregated licensed bandwidth is available for fixed communications. \rev{If these frequencies could be freed also for mobile communications, the larger channel bandwidths envisioned in Section III could be deployed, facilitating ultra-high bit rates and ultra-low latencies}. In general, one big question for beyond \SI{52.6}{GHz} communications is that how the {\color{black}overall available spectrum assets} can be efficiently shared between operators to allow ultra-high throughput operation and not to fragment the frequency bands into too small pieces.

\begin{table*}[]
    \setlength{\tabcolsep}{3pt}
    \renewcommand{\arraystretch}{1.3}
    \caption{Physical layer numerology for 5G NR Rel-15 and corresponding numerology considerations for beyond \SI{52.6}{GHz} communications. \rev{For reference, also an example of the 802.11ay numerology assuming normal guard interval and channel bonding of four \SI{2.16}{GHz} channels is shown.}}
    \centering
    \begin{tabular}{|l||c|c|c|c||c|c|c|c|c|c||c|c|}
        \hline
        \multirow{3}{*}{\textbf{Parameter}} &   \multicolumn{12}{c|}{\textbf{Value}} \\ \cline{2-13}
        &  \multicolumn{4}{c||}{\textbf{5G NR Rel-15}} & \multicolumn{6}{c||}{\textbf{Beyond \SI{52.6}{GHz} Evolution}} & \multicolumn{2}{c|}{\textbf{\rev{802.11ay}}} \\ 
        &  \multicolumn{4}{c||}{\textbf{DFT-s-OFDM \& OFDM}} & \multicolumn{6}{c||}{\textbf{DFT-s-OFDM \& OFDM}} & \textbf{\rev{SC}} & \textbf{\rev{OFDM}} \\
        \hline
        \hline
        SCS [kHz] & 15 &30 & 60  & 120 &\textbf{120} & 240  & 480 & 960 & 1920 & 3840 & \rev{3437.5} & \rev{5156.25} \\ 
        \hline
        Sampling freq. [MHz]  &   61.44  &122.88 &245.76 &491.52 &\textbf{491.52} &983.04 &1966.08 &3932.16 & 7864.32 & 15728.64 & \rev{7040} & \rev{10560} \\ \hline
        Slot duration [us] & 1000 & 500 &250 & 125  
        &\textbf{125} & 62.5  & 31.25 & 7.8125& 3.90625& 1.953125 & \multicolumn{2}{c|}{\rev{-}} \\ \hline
        FFT size  &   \multicolumn{4}{c||}{4096} &\multicolumn{6}{c||}{4096} & \multicolumn{2}{c|}{\rev{2048}} \\ 
        \hline
        Number of SCs per PRB & \multicolumn{4}{c||}{12} & \multicolumn{6}{c||}{12} & \multicolumn{2}{c|}{\rev{-}} \\ 
        \hline
        Max. number of PRBs &  270  &273 &264 &264 & \multicolumn{6}{c||}{264} & \multicolumn{2}{c|}{\rev{-}} \\
        \hline
       Max. allocation BW [MHz]  &   48.6    &98.28  &190.08 &380.16 &\textbf{380.16}   &760.32 &1520.64 &3041.28 &6082.56 &12165.12 & \rev{6160} & \rev{8306.7}\\ \hline
        Max. channel BW [MHz] &   50 &100 &200 &400    &\textbf{400} &800  &1600 &3200 &6400 & 12800 & \rev{8640} & \rev{8640}\\ 
        \hline
        PHY \rev{bit rate}, QPSK [Gb/s]   &   0.1     &0.2    &0.3    &0.6    &\textbf{0.6}     & 1.2   &2.4  & 4.8  & 9.6  & 19.2 & \rev{12.29} & \rev{13.27}\\ 
        \hline
        PHY \rev{bit rate}, 16-QAM [Gb/s]   &   0.2     &0.3    &0.6    &1.2    &\textbf{1.2}     & 2.4   &4.8  & 9.6  & 19.2  & 38.3 & \rev{24.57} & \rev{26.54}\\ 
        \hline
        PHY \rev{bit rate}, 64-QAM [Gb/s]   &   0.2     &0.5    &0.9    &1.8    &\textbf{1.8}     & 3.6   &7.2  & 14.4  & 28.8  & 57.5 & \rev{49.15} & \rev{53.08}\\ 
        \hline
        PHY \rev{bit rate}, 256-QAM [Gb/s]   &   0.3     &0.6    &1.2    &2.4    &\textbf{2.4}     & 4.8   &9.6  & 19.2  & 38.3  & 76.7 & \rev{-} & \rev{-} \\ 
        \hline
    \end{tabular}
    \label{tab:PHY_layer_parameterization}
\end{table*}

\vspace{0mm}
\section{5G NR Physical Layer: \\ Current Status and Beyond \SI{52.6}{GHz} Evolution}
\label{sec:5GNowAndBeyond52p6GHz}
\rev{\subsection{Physical Layer Numerology of 5G NR Rel-15} }
The 5G NR \rev{Rel-15} was designed to support wide range of SCSs to handle different uses cases and a wide range of supported carrier frequencies. In Rel-15, two frequency ranges are defined, where also different SCSs are supported \cite{2017:J:Parkvall:5GNR}. The supported SCSs follow the scaling of \SI{15}{kHz} with powers of two, defined as $15 \times 2^\mu$~kHz, where $\mu \in \{0,1,2,3,4\}$, corresponding to SCSs of 15/30/60/120/240~kHz, respectively. The frequency range 1 (FR1) is defined for carrier frequencies \SI{410}{MHz} -- \SI{7.125}{GHz} and supports SCSs of 15/30/60~kHz, while frequency range 2 (FR2) defined for frequency range \SI{24.25}{GHz} -- \SI{52.6}{GHz} supports 60/120/240~kHz SCSs, where \SI{240}{kHz} SCS is only allowed for the so-called synchronization signal block \cite{3GPPTS38300}. The \rev{5G NR Rel-15} related main PHY parameters are summarized in Table \ref{tab:PHY_layer_parameterization}. It is also reminded that in 5G NR numerology, the time duration of the slot and the cyclic prefix (CP) is decreased with increasing SCS and sampling frequency. \rev{Finally, Rel-15 supports OFDM in DL while both OFDM and DFT-s-OFDM in UL}.  

\vspace{-3mm}
\rev{\subsection{Physical Layer Numerology for Beyond \SI{52.6}{GHz} Evolution}}
The current Rel-15 specification does not provide sufficient flexibility and power efficiency for communications above \SI{52.6}{GHz} carrier frequencies while achieving multi-gigabit throughputs. 
\rev{Specifically, in beyond \SI{52.6}{GHz} communications, higher SCSs may be required -- due to the increased PN distortion as well as increased Doppler frequencies, which can be both partially mitigated through shorter multicarrier symbols}. As the PN estimation and compensation based on PTRS is one of the main themes of this paper, it is thoroughly discussed in Section \ref{sec:pn_models_and_PTRS} together with considered PN models. \rev{Additional important reason for increased SCS is the capability to achieve extremely high channel bandwidths with reasonable FFT size, as required in beyond 52.6GHz}. 
Increasing the SCS leads also to reduced PHY latency, but it may also lead to some system design difficulties. 

\rev{To address the issues discussed above,} the \rev{most important} component to \rev{modify} is the basic PHY numerology, as shown in Table \ref{tab:PHY_layer_parameterization}. Inherited from the Rel-15 NR, we assume that the supported SCSs in beyond \SI{52.6}{GHz} still follow the scaling of \SI{15}{kHz} SCS with powers of two, as shown in Table \ref{tab:PHY_layer_parameterization}. Similar to Rel-15, we assume that FFT size of 4096 samples is used as a baseline, and that the maximum number of physical resource blocks (PRBs) equals 264, as currently defined in \cite{3GPPTS38101-2} for FR2 with \SI{60}{kHz} and \SI{120}{kHz} SCSs. By allocating 180 PRBs with \SI{960}{kHz} SCS, the allocation bandwidth corresponds to \SI{2.07}{GHz} which is well suited to the \SI{2.16}{GHz} channel bandwidth, corresponding to the IEEE wireless local area network (WLAN) 802.11ay channel spacing \cite{2018:J:Zhou:80211ayTutorial}. 

The achievable PHY \rev{bit rates} with maximum allocation bandwidths and different modulations are shown in Table~\ref{tab:PHY_layer_parameterization}. The \rev{bit rates} are obtained by considering a rank-1 transmission with a slot of 14 OFDM symbols, from which one symbol is reserved for physical downlink control channel (PDCCH) and one for demodulation reference signal (DMRS) used for channel estimation. In addition, PTRS overhead of 48 subcarriers corresponding to some $1.5\%$ is assumed. This example shows that to reach larger than \SI{10}{Gb/s} PHY \rev{bit rate}, at least \SI{2}{GHz} of bandwidth per operator should be considered. \rev{We also note that the fundamental numerologies and the PHY bit rates are the same for OFDM and DFT-s-OFDM.}

\rev{Even though the increased subcarrier spacing is one of the most important components in the system design}, use of lower SCS can also be desirable for several reasons. Firstly, it allows supporting longer CP length in time domain, alleviating synchronization and beam switching procedures. Secondly, it provides higher power spectral density (PSD) for transmitted signals with equal number of subcarriers. Thirdly, it decreases the sampling rates required by the UE, thus enabling reduced power consumption and higher coverage for beyond \SI{52.6}{GHz} communications. \rev{Furthermore, use of lower SCS enables support for users with lower bandwidth capabilities.} Moreover, increasing the supported channel bandwidth with increased SCSs leads to increased transmitter (TX) \rev{distortion} and RX noise power which limits the coverage of the system. 

\vspace{-3mm}
\rev{\subsection{Relation to IEEE WLAN 802.11ay}} 
\vspace{-2mm}
\rev{The IEEE WLAN 802.11ay, and its predecessor 802.11ad, are important references for wireless communications at \SI{60}{GHz} band \cite{2017:J:Ghasempour:80211ay,2018:J:Zhou:80211ayTutorial}. Although the system requirements, such as coverage area and mobility support, are completely different, the need for power efficient physical layer is of common interest. In 802.11ay, the single-carrier (SC) support is mandatory while the OFDM support is optional. This is due to the uncertainty on the benefits of OFDM in local area communications where the radio link distances are clearly smaller than what is envisioned by 3GPP for mobile access. It is also noted that in 802.11ay, the SC mode assumes utilizing a known Golay sequence as a guard interval between SC data symbol blocks, where the total length of data symbol block together with the guard interval corresponds to the used FFT size. Thus, 802.11ay SC can be considered as a unique word DFT-s-OFDM, instead of the cyclic prefix approach of 5G NR.}

\rev{The physical layer of 802.11ay is clearly simpler than 5G NR, with only a single SCS being supported for SC or OFDM mode, limited number of supported modulation and coding schemes, and clearly smaller requirements on the link reliability. In 802.11ay, FFT size of 512 is used to support channel bandwidth of \SI{2.16}{GHz}, which can be extended with channel bonding up to maximum contiguous bandwidth of \SI{8.64}{GHz}.}
\rev{Channel bonding relates to operation where the used FFT size is increased to allow the use of larger channel bandwidth with the fixed SCS. Therefore, for 5G NR evolution to provide comparable instantaneous PHY bit rates with WLAN 802.11ay technology, channel bandwidths up to \SI{8.64}{GHz} should be supported. For reference, the basic 802.11ay numerology is also shown in Table~\ref{tab:PHY_layer_parameterization}. For the rank-1 PHY bit rates, we have assumed a continuous transmission burst of \SI{2}{ms}, corresponding to the maximum physical layer convergence protocol data unit duration, and included the overhead of different training or channel estimation fields, headers, and interframe space between bursts. Considering the allocated bandwidth, the theoretical bit rate of 802.11ay is larger due to smaller control and training overhead with large transmission bursts, but when comparing with respect to the channel bandwidth, 5G NR provides larger PHY bit rate due to significantly better spectrum utilization. }

\vspace{5mm}
\section{Phase Noise Challenge and PTRS Designs}
\label{sec:pn_models_and_PTRS}

\subsection{Phase Noise Fundamentals}
\label{subsec:pn_models}

Phase noise is typically characterized through a PSD mask, where the noise power within a \SI{1}{Hz} bandwidth at a certain frequency offset from the carrier frequency is defined, relative to the noise power at the carrier frequency. Generally, the higher is the offset, the lower is the PSD response of the PN \cite{2000:J:Lee:OscillatorTutorial}. Since OFDM and DFT-s-OFDM use multiple orthogonal subcarriers transmitted at different center frequencies, they are both affected quite similarly under PN. More specifically, PN causes a common phase error (CPE) which affects all the subcarriers within a multicarrier symbol similarly \cite{2019:L:Syrjala:blockPTRS}. This means that only a single complex value is required to compensate this term from the received signal. However, due to the relatively wide PSD response of the PN in mmWave communications, it also \rev{causes} inter-carrier interference (ICI). This effect can be mitigated by increasing the SCS, or by applying PTRS designs which allow for the estimation and compensation of the ICI components. Therefore, SCS is an important design parameter and the higher is the assumed center frequency, the higher is the required SCS, typically.

There are different PN models defined in 3GPP \cite{3GPPTR38803}, and PN modeling has a significant effect on the radio link performance. The 3GPP models are based on extensive studies on current trends in phase-locked loops (PLLs) and are publicly available. In general, there are two different local oscillator (LO) strategies for carrier frequency generation. The first strategy is based on a centralized LO where a single PLL is \rev{shared by all the RF transceivers}, while the second strategy is based on distributed carrier generation with individual PLL per each \rev{RF transceiver}. All the evaluations in this paper are based on the first strategy, i.e., we assume that there is only one PLL shared by all involved transceivers. \rev{This is a practical but also the worst-case assumption from the performance point of view, because distributed carrier generation would give some phase noise averaging gain when the signals are combined in the receiver from different antenna ports.} 

In our later numerical results, we use the \rev{centralized LO} PN model defined in \cite[Section 6.1.11]{3GPPTR38803}, which considers complementary metal oxide semiconductor (CMOS) based design for the UE due to lower cost and power consumption, and Gallium Arsenide (GaAs) based design for the BS. The GaAs based oscillators are more expensive and not as suitable to highly integrate circuits, and therefore not as well suited for UEs as CMOS based designs. The power consumption of the UE model is set to \SI{50}{mW} and for the BS it is set to \SI{80}{mW}, and the loop bandwidth for the PLL-based PN models is \SI{187}{kHz} for the UE model and \SI{112}{kHz} for the BS model.

\begin{figure}
    \centering
    \includegraphics[width=0.9\columnwidth]{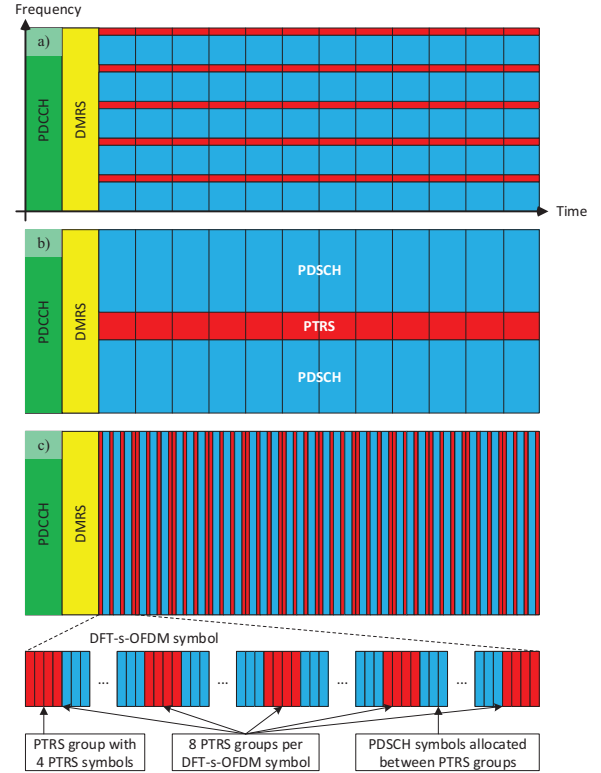}
    \caption{Illustration of Rel-15 NR PTRS structures for (a) OFDM and (c) DFT-s-OFDM. In addition, the considered novel block PTRS structure for OFDM is illustrated in (b).}
    \label{fig:PTRS_structure}
\end{figure}

{
\floatstyle{boxed} 
\restylefloat{figure}
\begin{figure*}[ht]
  \centering
  \vspace{-1mm}
  \subfloat[][]{\includegraphics[angle=0,width=0.96\columnwidth]{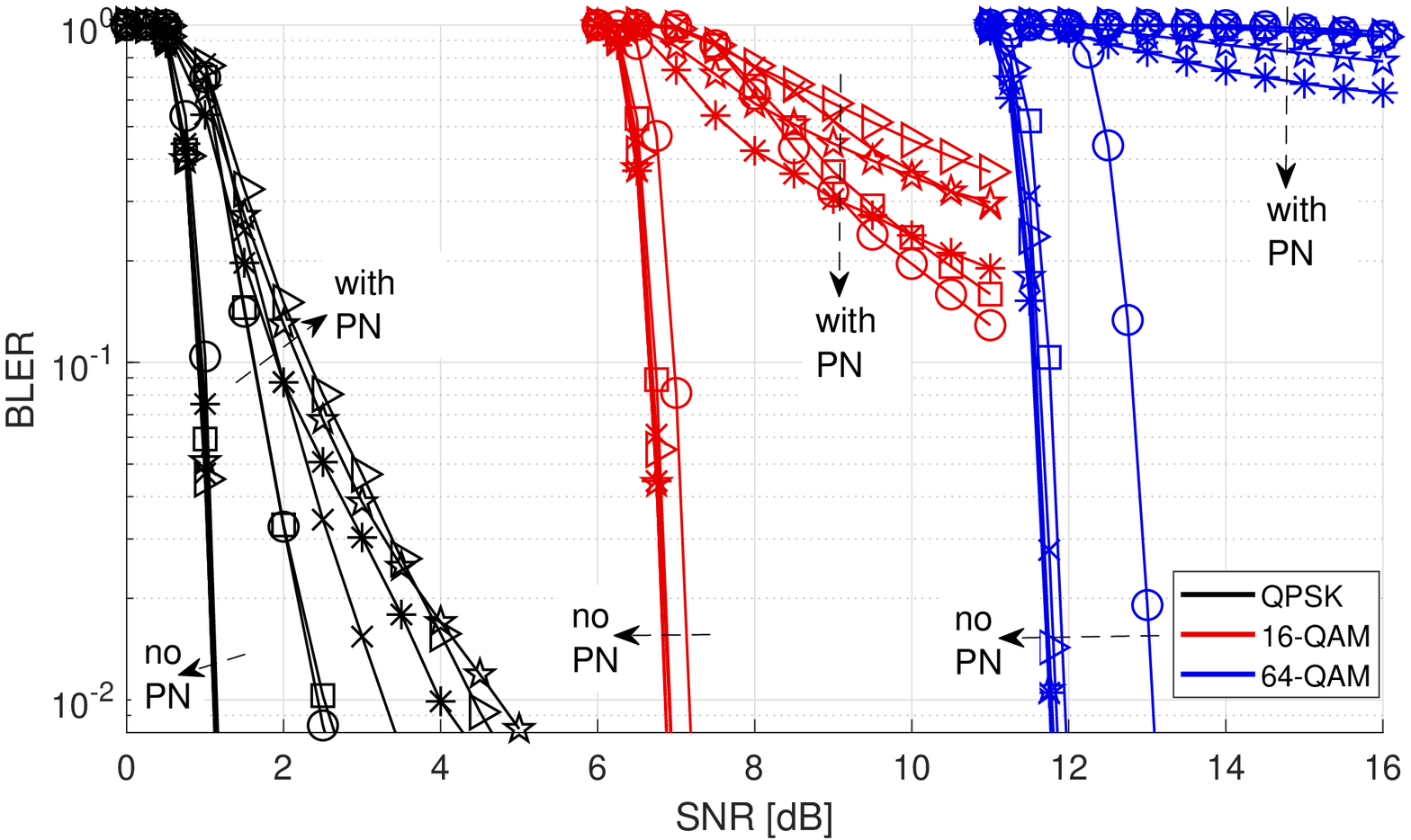}}  
  \qquad
  \subfloat[][]{\includegraphics[angle=0,width=0.96\columnwidth]{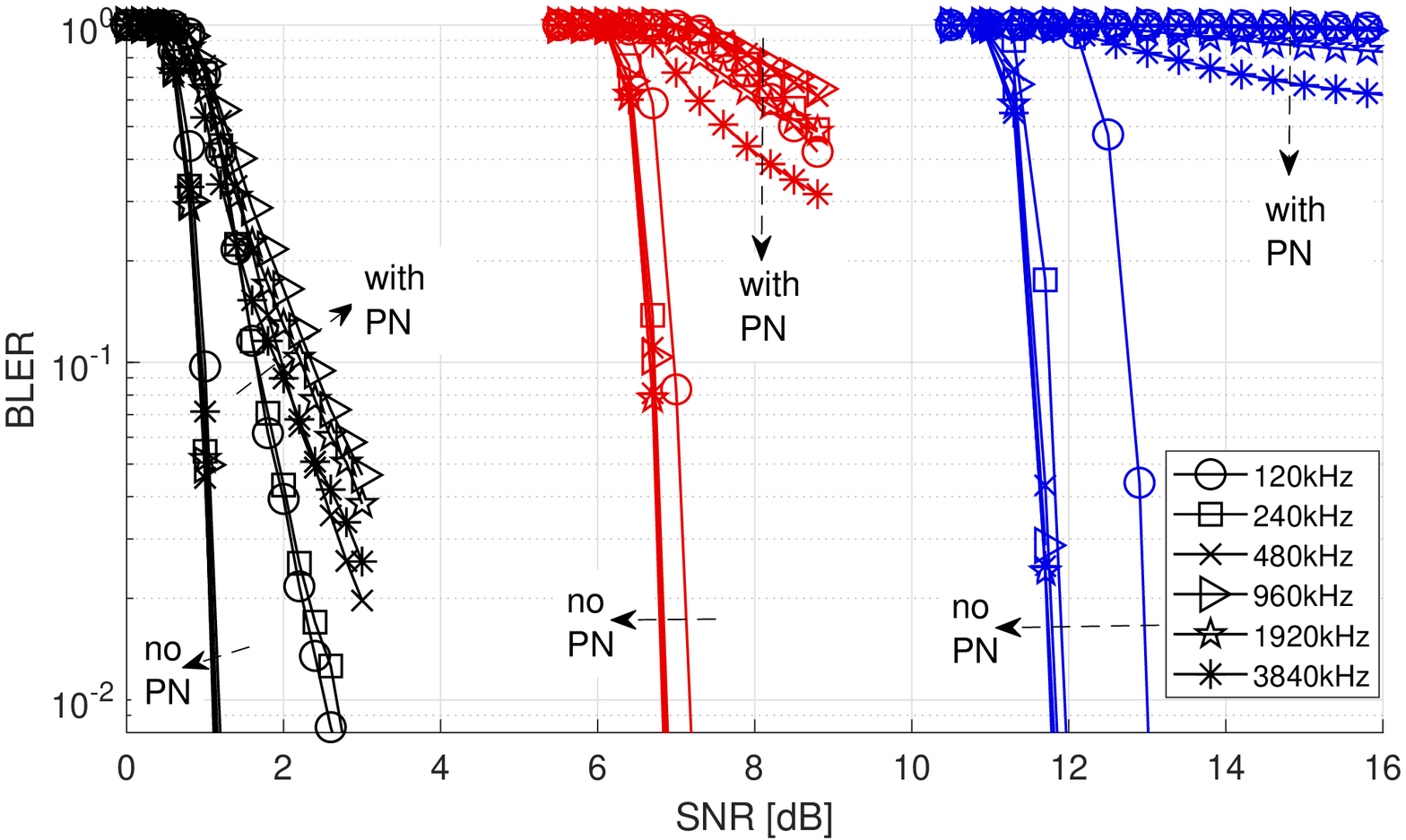}}
  \vskip\baselineskip
  \vspace{-4mm}
  \caption{\rev{Example radio link performance with and without PN when using (a) OFDM (b) DFT-s-OFDM. No PTRS based compensation is yet considered.}}
  \label{fig:60GHz_NoPTRS}
\end{figure*}
}

\subsection{PTRS Designs of Rel-15 5G NR}
\label{subsec:rel15_ptrs}

\subsubsection{Design for OFDM}
\label{subsubsec:Rel15PTRSOFDM}

In the case of OFDM signal, individual PTRS symbols are inserted in the frequency domain with predefined frequency gap, as illustrated in Fig \ref{fig:PTRS_structure} (a), where the physical downlink shared channel (PDSCH) carries the user data in DL direction. Thus, the PTRS structure for OFDM relies on so-called distributed PTRS design, occupying individual subcarriers with predefined distance in frequency. Rel-15 supports inserting PTRS to every second or fourth PRB in frequency domain. Since PN varies rapidly over time, PTRSs need to be inserted densely in time. Therefore, every $L$th OFDM symbol in time domain, where $L \in \{1,2,4\}$, can contain a PTRS. In the numerical evaluations, we assume the maximum density for Rel-15 PTRS which leads to \rev{overhead of $1/(2\times12)=4.2\%$}. Distributed frequency-domain insertion means that only CPE can be accurately estimated and compensated for each OFDM symbol containing PTRS, which significantly limits the performance with lower SCS or high order modulations, as will be shown in Section \ref{sec:results}. 

In the RX, after channel estimation and equalization procedures, one can calculate the rotation of each PTRS in each OFDM symbol and take the average of these to obtain CPE estimate, and finally compensate it before detection and decoding procedures. In the case of not inserting PTRS to each OFDM symbol, CPE estimates for those OFDM symbols without PTRS are obtained by interpolating the available PTRS estimates in the time domain, leading however to increased detection latency and buffering requirements.

\subsubsection{Design for DFT-s-OFDM}

In the Rel-15 5G NR standardization phase, two different methods to insert PTRSs for DFT-s-OFDM signal were considered: pre-DFT and post-DFT insertion. That is, inserting PTRS either in time domain or frequency domain. The latter one basically would enable exactly the same compensation mechanisms as with OFDM. However, the former one was accepted to specifications due to its lower PAPR behaviour and better PN compensation capabilities. More specifically, reference symbols are inserted before DFT to enable sample-level time domain PN tracking. 

The Rel-15 NR defines different configurations for group-based time domain PTRS, where either 2 or 4 samples per group are used, and 2, 4, or 8 groups per DFT-s-OFDM symbol are supported \cite[Table 6.4.1.2.2.2-1]{3GPPTS38211}. Thus, the maximum number of \rev{PTRS resources} per DFT-s-OFDM symbol is $8 \times 4 = 32$, which results in \rev{overhead of $1.5\%$ per DFT-s-OFDM symbol when $12 \times 180 = 2160$}  
subcarriers are used. This configuration is used as a Rel-15 baseline in Section \ref{sec:results}.  

The high level concept of DFT-s-OFDM PTRS is illustrated in Fig. \ref{fig:PTRS_structure} (c) together with DFT-s-OFDM symbol wise PTRS allocation assuming maximum PTRS configuration. Due to distributing the PTRS symbols in the time domain, the RX can track the time varying PN within each DFT-s-OFDM symbol. In the RX, after the frequency-domain channel equalization, the received DFT-s-OFDM signal is converted back to time domain using inverse DFT, after which the PN can be estimated from the time domain PTRS and compensated before detection and decoding procedures. For example, one can calculate the mean rotation in each PTRS group and use a simple linear interpolation to get the estimated PN values between the time domain PTRS groups. Note that with DFT-s-OFDM, the PTRS design allows a computationally efficient implementation to track and compensate time-varying PN response within a DFT-s-OFDM symbol, which is not possible with the Rel-15 NR distributed PTRS design for OFDM.

{
\floatstyle{boxed} 
\restylefloat{figure}
\begin{figure*}[t]
  \centering
  \vspace{-2mm}
  \subfloat[][]{\includegraphics[angle=0,width=0.82\columnwidth]{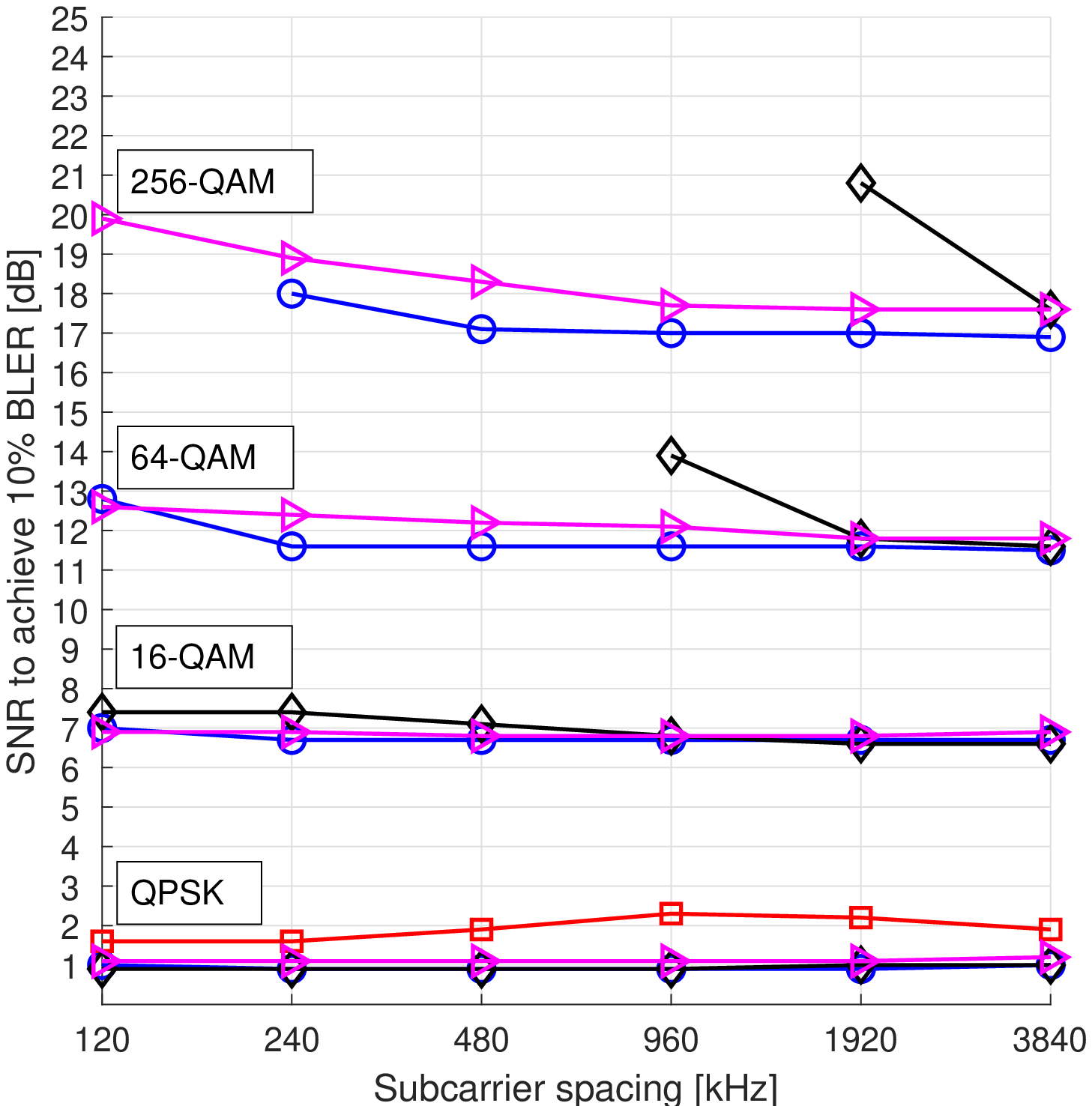}}
  \qquad
  \hspace{2mm}
  \subfloat[][]{\includegraphics[angle=0,width=0.82\columnwidth]{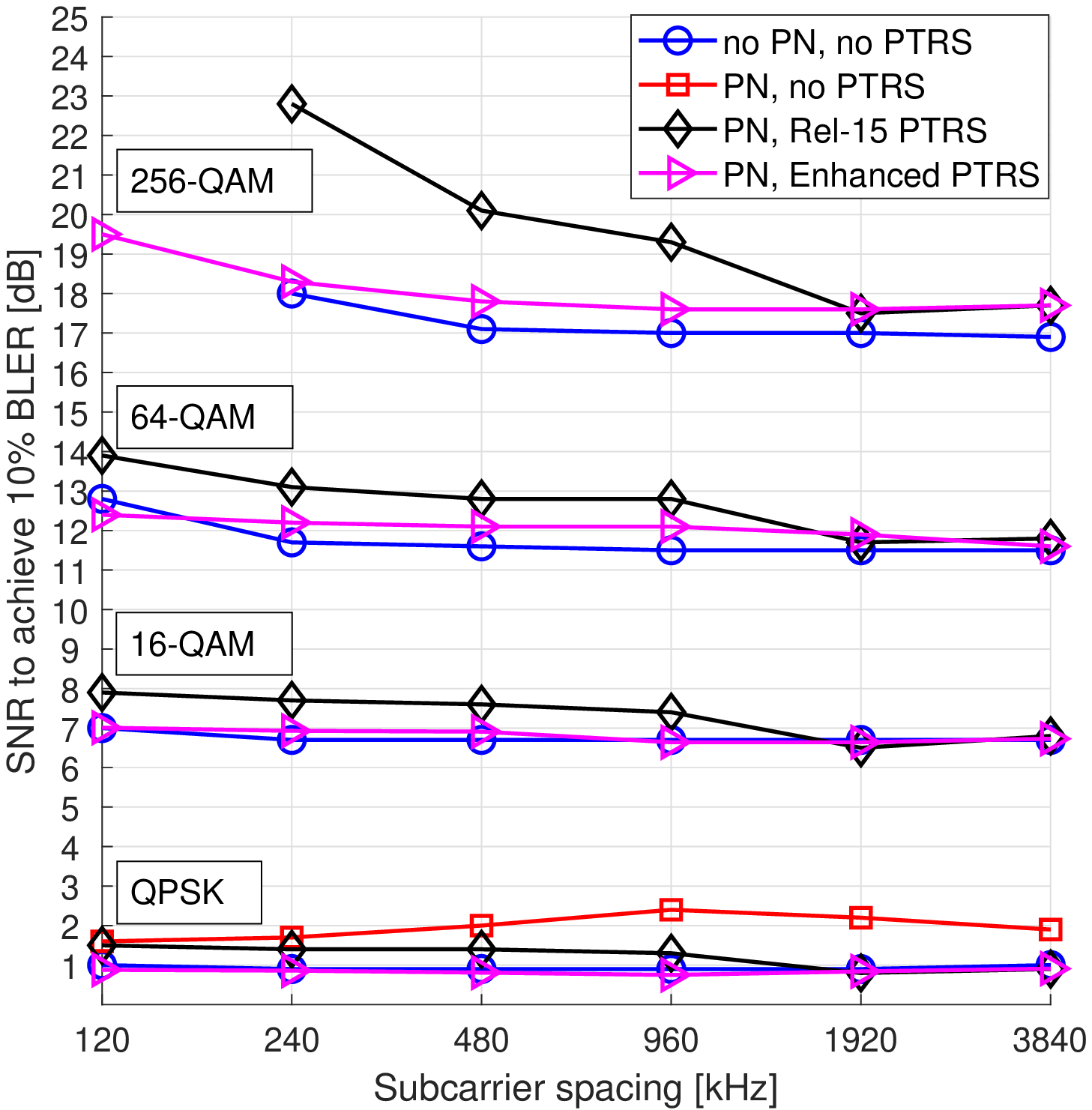}}
  \vskip\baselineskip
  \vspace{-4mm}
  \caption{\rev{Required SNR to achieve 10\% block-error rate target at \SI{60}{GHz} with different subcarrier spacings for a) OFDM and b) DFT-s-OFDM.} }
  \label{fig:60GHz_WithPTRS}
\end{figure*}
}

\vspace{-1mm}
\subsection{New PTRS Designs for Beyond \SI{52.6}{GHz} Communications}
\subsubsection{Block PTRS Design for OFDM}
The concept of frequency domain block PTRS is introduced in \cite{2019:L:Syrjala:blockPTRS}. The basic idea is to allocate a frequency contiguous block of PTRS symbols, as shown in Fig. \ref{fig:PTRS_structure} (b), which allows to estimate PN induced frequency-domain ICI components at RX. 
As the current Rel-15 specification dictates a specific frequency resolution for distributed PTRS, it is possible that with block PTRS based design one can achieve better performance with lower reference signal overhead in wide channels using fullband allocations. Typically, it is considered that the block PTRS would be allocated as multiples of PRBs, where each PRB contains 12 subcarriers, to simplify control. However, block PTRS can also be allocated even with subcarrier resolution to maximize spectral efficiency, as long as the used block size is equal to or larger than the number of unknowns in the estimation process \cite{2019:L:Syrjala:blockPTRS}. Block PTRS is inserted to each OFDM symbol, as the time continuity of ICI components is typically not guaranteed, and thus interpolation is not possible. On the other hand, having block PTRS in each OFDM symbol supports highly efficient pipelined RX architecture.

\rev{In addition to PN induced ICI, block PTRS allows to some extent compensate also for the ICI induced by time-varying channel (Doppler) and is thus well suited also for high-mobility communications where the residual Doppler effects might be significant.} Also in low-mobility scenarios in beyond \SI{52.6}{GHz} communications, as will be shown in Section \ref{sec:results}, block PTRS allows to improve the link performance with front-loaded designs (i.e., a single DMRS in the beginning of the slot), as the time-varying channel during the slot duration causes ICI which is then mitigated with the block PTRS design and related compensation algorithms. In the numerical evaluations a block PTRS of size 4 PRBs, or 48 subcarriers, is assumed leading to \rev{overhead of $2.2\%$ when assuming $12\times180=2160$} active subcarriers, which is clearly less than with the Rel-15 NR PTRS design.

\subsubsection{PTRS Design Enhancements for DFT-s-OFDM}

For beyond \SI{52.6}{GHz} communications, it is important to study whether the Rel-15 NR maximum PTRS configuration is sufficient to tackle the increasing PN in the higher frequencies, or can we obtain significant performance improvements by defining new configurations. In order to improve the PN estimation capability with DFT-s-OFDM there are basically two options: 1) increasing the number of PTRS symbols per group, and 2) increasing the number of PTRS groups within the DFT-s-OFDM symbol. Increasing the number of PTRS symbols per group basically provides averaging gain against noise and interference, and does not directly improve our capability to estimate fastly changing PN response. Therefore, our proposal for the enhanced PTRS design for DFT-s-OFDM focuses on increasing the number of PTRS groups, to allow improved PN response tracking within the DFT-s-OFDM symbol. The detailed evaluation for optimized design is outside the article scope, and thus a design leading to the same overhead as the block PTRS proposed for OFDM is selected. Thus, we propose a new PTRS configuration for DFT-s-OFDM using 12 PTRS groups with four PTRS symbols per group, which results in \rev{$2.2\%$ overhead when assuming 2160} 
active subcarriers, corresponding to the OFDM block PTRS overhead.

\vspace{-1mm}
\section{Radio Link Performance at \SI{60}{GHz}}
\label{sec:results}
In this section, the performance of OFDM and DFT-s-OFDM waveforms with or without PTRS and with or without PN induced interference is evaluated over the selected SCS values shown in Table \ref{tab:PHY_layer_parameterization} and discussed in Section \ref{sec:5GNowAndBeyond52p6GHz}. We assume a maximum channel bandwidth of \SI{2.16}{GHz}, which follows the channelization for WLAN 802.11ay operating in \SI{60}{GHz} unlicensed band \cite{2018:J:Zhou:80211ayTutorial}. Therefore, the maximum allocation size is limited to 180, 90, or 45 PRBs with SCS \SI{960}{kHz}, \SI{1920}{kHz}, or \SI{3840}{kHz}, respectively. To obtain comparable performance with smaller SCS, we have limited the maximum allocation size to 180 PRBs also with SCSs less than \SI{960}{kHz}.

 The evaluated PN model is described in Section \ref{subsec:pn_models}, that is, it follows BS and UE models for TX and RX, respectively, as defined in \cite[Section 6.1.11]{3GPPTR38803}. The evaluations concentrate on the DL rank-1 transmission scheme with polarization specific antenna panels in BS and UE. There are 128 antenna elements organized into a 8x16 antenna array per polarization at the BS and 16 antenna elements organized into a 4x4 antenna array per polarization at the UE. The link performance is evaluated with QPSK, 16-QAM, 64-QAM, and 256-QAM modulations with fixed coding rate of $R=2/3$. The used channel codec is a 5G NR Rel-15 compliant LDPC code. The used channel model is clustered delay line E (CDL-E) with \SI{10}{ns} root-mean-squared (RMS) delay spread and Rician factor $K=$\SI{15}{dB} \cite{3GPPTR38901}\rev{\footnote{\rev{Directly available through Mathworks nrCDLChannel System object$^{\textrm{TM}}$.}}}. In all cases, \SI{3}{km/h} UE mobility is assumed.

\subsection{Radio Link Performance Without PTRS}

First, to illustrate the significant effect of PN in beyond \SI{52.6}{GHz} communications, the performance without PTRS and with or without PN is shown in Fig. \ref{fig:60GHz_NoPTRS}, for selected modulations. \rev{From the PN-free results, we can observe that all the SCSs provide similar link performance for both waveforms. With the assumption of a slot based transmission with only 1~DMRS symbol, the channel variation caused by Doppler during the 14~OFDM symbol slot starts to degrade the link performance of 64-QAM with \SI{120}{kHz} SCS. As will be observed later on, ICI compensation allows to alleviate this problem.} 
When considering the effect of PN, we note that QPSK modulation can be supported without PTRS with approximately \SI{1}{dB}-\SI{2}{dB} loss in the required SNR. In addition, 16-QAM could be used without PTRS at \SI{60}{GHz} carrier frequency, but with significant loss in the required SNR to achieve $10\%$~block error rate (BLER), indicating the need for PN estimation and compensation already with low-order modulations. High-order modulations, 64-QAM and 256-QAM, do not work at all without PTRS under PN, as is observed also from Fig. \ref{fig:60GHz_WithPTRS}. \rev{Full BLER curves for all modulation orders are available in the supplementary electronic materials.}

\subsection{Radio Link Performance With PTRS}

The \rev{Fig. \ref{fig:60GHz_WithPTRS} compares the link performance when the Rel-15 or enhanced PTRSs are used to compensate the PN. The performance with or without PN and without PTRS are also provided for reference. The x-axis includes different subcarrier spacings, while the y-axis shows the required SNR to achieve 10\%~BLER, which is used as a typical operating point in adaptive modulation and coding for the first transmission.} Throughout this section, in the case of DFT-s-OFDM, we use either the time domain CPE estimate or the interpolated PN estimate, depending on which gives the best result. With OFDM and Rel-15 PTRS designs, only CPE compensation is possible. In the case of block PTRS \cite{2019:L:Syrjala:blockPTRS}, it is assumed that a contiguous allocation of 4~PRBs at the middle of the band is used, and on the RX side four PN frequency components from both sides of the DC are estimated and used in the PN compensation. It should be noted that the block PTRS based design simultaneously estimates both the CPE and the ICI components.

\rev{We can first }conclude that the performance of lower-order modulations, QPSK and 16-QAM, can be clearly improved with Rel-15 PTRS, although these are typically assumed to operate without PTRS. \rev{For QPSK, the use of PTRS gives approximately \SI{1}{dB}-\SI{2}{dB} gain in the SNR axis.} In addition, Fig. \ref{fig:60GHz_WithPTRS} (a) shows that Rel-15 PTRS design for OFDM can support 64-QAM if SCS $\geq$ \SI{960}{kHz} is used. On the other hand, as shown in Fig. \ref{fig:60GHz_WithPTRS} (b), DFT-s-OFDM with Rel-15 PTRS design performs significantly better with 64-QAM, allowing to use all considered SCSs. From Fig. \ref{fig:60GHz_WithPTRS} (a), we also note that OFDM with Rel-15 PTRS can support 256-QAM with SCS $\geq$ \SI{1920}{kHz}, whereas DFT-s-OFDM with Rel-15 PTRS can support 256-QAM already with SCS \SI{240}{kHz}. Nevertheless, it is clear that if Rel-15 PTRS designs are not updated for OFDM or DFT-s-OFDM, significant radio link performance degradation is expected with the largest currently supported SCS of \SI{120}{kHz}. \rev{The results shown in Fig. \ref{fig:60GHz_WithPTRS} (a) and (b) indicate, that by directly extending 5G NR Rel-15 FR2 operation to \SI{60}{GHz} carrier frequency, data modulations only up to 16-QAM modulation can be supported by OFDM waveform, where as DFT-s-OFDM based downlink for user data could support modulation orders up to 64-QAM.}

On the other hand, Fig. \ref{fig:60GHz_WithPTRS} (a) shows that by using the novel block PTRS with OFDM improves the performance significantly when compared to Rel-15 results, and even 256-QAM can be supported with all evaluated SCSs. This highlights the need for block PTRS support with OFDM in beyond \SI{52.6}{GHz} communications, especially if considering the extension of 5G NR Rel-15 FR2 operation to \SI{60}{GHz} carrier frequency, and also the dominance of PTRS designs allowing ICI compensation over CPE compensation. For DFT-s-OFDM, increasing the number of PTRS groups from 8 to 12 can improve the performance significantly. More specifically, the used 12 PTRS groups allow to support 256-QAM over all evaluated SCS with even better link performance than OFDM using block PTRS. Thus, DFT-s-OFDM based downlink for user data combined with new PTRS configurations would allow to support 256-QAM modulation if current Rel-15 NR FR2 operation is extended to \SI{60}{GHz} carrier frequency. 

\rev{Related to the Doppler induced ICI with \SI{120}{kHz} SCS with front-loaded DMRS design, it can be observed from Fig. \ref{fig:60GHz_WithPTRS} that with both waveforms the radio link performance degrades with 64-QAM, while 256-QAM does not anymore reach 10\%~BLER. With the enhanced PTRS designs, we can compensate also this distortion, allowing us to improve 64-QAM performance and enable 256-QAM performance with \SI{120}{kHz} SCS. In addition to significant improvements with high-order modulations, also with QPSK and 16-QAM clear link performance improvements are observed with enhanced PTRS designs with both waveforms. In general, increasing SCS allows to tolerate more Doppler spread and due to shorter OFDM symbols, the tracking of time variation of the channel is easier with fixed reference signal overhead.}

\begin{figure}
    \centering
    \vspace{-5mm}
    \includegraphics[width=0.95\columnwidth]{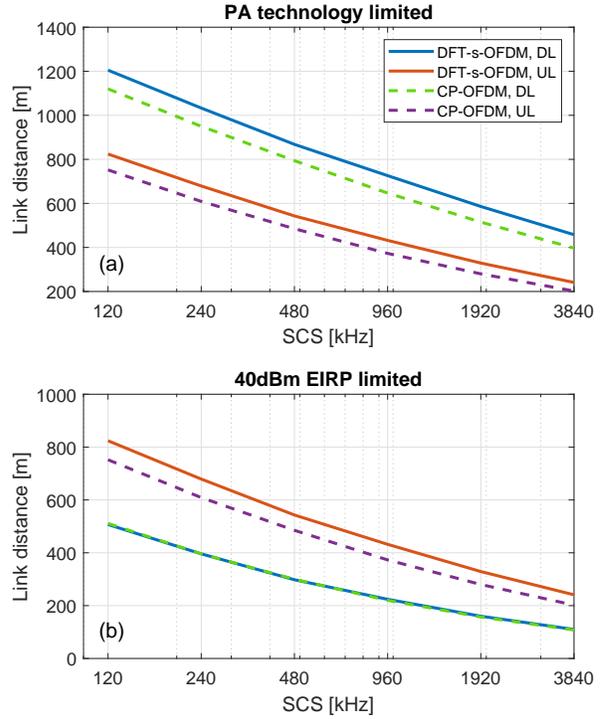}
    \vspace{-3mm}
    \caption{\rev{Illustration of the achievable LOS link distance with CP-OFDM and DFT-s-OFDM assuming 16-QAM modulation in a) PA technology limited scenario and b) \SI{40}{dB} EIRP limited scenario.}}
    \vspace{3mm}
    \label{fig:link_distance}
\end{figure}

\rev{\section{Link Budget Considerations for Beyond \SI{52.6}{GHz} Communications}}
\label{sec:linkBudget}
We finally consider link budget aspects for beyond \SI{52.6}{GHz} communications, via example evaluations with the provided link budget tool that is available through the electronic materials. \rev{The UL and DL link distances provided by the link budget tool are based on urban micro (UMi) LOS and UMi non-line-of-sight (NLOS) PL models \cite[Table 7.4.1-1]{3GPPTR38901}, while also consider the effect of atmospheric gas attenuation. Furthermore, we evaluate two different scenarios, EIRP limited and PA technology limited scenario when estimating the link distances. 

Because the beyond \SI{52.6}{GHz} communications is assumed to be mainly time division duplex-based, we consider only full allocations for both DL and UL. In addition, 32~antenna elements for the UE and 256~antenna elements for the BS beamforming antenna arrays are assumed. 16-QAM modulation is considered, and the required signal-to-noise ratio (SNR) values are based on the results presented in Section \ref{sec:results}. In the RX side, we consider a linearly increasing noise figure with respect to the used bandwidth to include the possible effects of TX and RX non-idealities, which tend to increase with supported bandwidth \cite{3GPPTS38101-2}. }

\rev{Based on the overview of different PA output saturation powers presented in \cite{2014:J:Wu:PAoutputPowerWband}, we evaluated the arithmetic mean of these values corresponding to \SI{15.4}{dBm}, and use this as an approximative estimate for the PA saturation power available in BS and UE. Then considering the maximum power reduction required by 16-QAM modulation, we have used the output power backoff of \SI{4.5}{dB} for DFT-s-OFDM and \SI{6.5}{dB} for CP-OFDM, as defined in \cite[Table 6.2.2.1-2]{3GPPTS38101-2}. This leads to the maximum PA output power per antenna element to be \SI{10.9}{dBm} with DFT-s-OFDM and \SI{8.9}{dB} with CP-OFDM. Because CP-OFDM requires a larger maximum power reduction, it provides a reduced maximum PA output power and link distance when compared to DFT-s-OFDM. In general, allowing the use of DFT-s-OFDM in DL would improve the power efficiency of the BS because less power backoff is required and it could also enable the use of more non-linear PAs, leading to reduced cost of BS equipment. Furthermore, using DFT-s-OFDM in downlink would enable the use of new pulse shaped Pi/2-BPSK modulation defined in 5G NR Rel-15 for further coverage extension. }

\rev{Traditionally, the link distance is mobile communications is limited by the UL. This is observed also in Fig. \ref{fig:link_distance} (a), where the link distance is limited by the assumptions made on the PA output power and waveform specific maximum output power reduction. As can be seen, link distance in DL is clearly larger due to larger EIRP, while DFT-s-OFDM provides some \SI{8}{\%} and \SI{10}{\%} larger link distance in DL and UL, respectively. This gain is mainly due to the fixed PA technology assumption, which allows DFT-s-OFDM to achieve larger output power and better power efficiency, compared to OFDM. }

\rev{However, in beyond \SI{52.6}{GHz} communications the UL vs. DL situation might be drastically different as in some frequency bands the same EIRP limit applies to both ends, the BS and the UE. In Fig. \ref{fig:link_distance} (b), the EIRP is limited to \SI{40}{dBm} following the current regulation in Europe regarding unlicensed communications at \SI{57}{GHz} - \SI{66}{GHz} band \cite[Table 4.2.2.1-2]{3GPPTR38807}. This leads to a scenario, where maximum EIRP in BS TX (corresponding to DL scenario) is achieved with smaller PA output power and therefore CP-OFDM and DFT-s-OFDM provide the same link distance results. On the other hand, due to the larger RX beamforming gain in the BS, the UL link distance is clearly larger, and {dominated by the DFT-s-OFDM based waveform}. This sort of imbalance between DL and UL in EIRP limited scenarios is often ignored in the system design or evaluation studies, and is thus here highlighted. It is also noted, that due to the given assumptions, the UL results for CP-OFDM and DFT-s-OFDM are identical in the EIRP limited and PA technology limited scenarios.}

\vspace{5mm}
\section{Conclusions and Future Perspective}
\label{sec:conclusions}
In this paper, the evolution of the 5G NR technology to the beyond \SI{52.6}{GHz} bands was discussed, with specific emphasis on the physical layer numerology, the phase noise challenge and the comparison of OFDM and DFT-s-OFDM waveforms. Specifically, the performance and limitations of the current PTRS designs of 5G NR Rel-15 standard were evaluated and compared to the enhanced PTRS structures, which include a novel block PTRS design for OFDM waveform and improved PTRS configurations for DFT-s-OFDM waveform. 

With the extensive set of performance evaluations, it was demonstrated that the DFT-s-OFDM performs significantly better than OFDM when using Rel-15 PTRS structures, since it allows to estimate the time varying PN response within the DFT-s-OFDM symbol. Furthermore, it was shown that even QPSK performance can be improved using PTRS, and that the proposed enhanced PTRS designs can support up to 256-QAM with all the evaluated subcarrier spacings up to \SI{3840}MHz. Therefore, the existing Rel-15 FR2 solutions can be extended to \SI{60}{GHz} carrier frequency if a new block PTRS design is introduced for OFDM, or DFT-s-OFDM based downlink with new PTRS configurations is introduced for user data, \rev{while substantially larger channel bandwidths and thus larger bit rates and reduced latencies are available through the new increased SCSs}.  

In the long run, a single solution is preferred for mobile communications over the whole frequency range of \SI{52.6}{GHz} - \SI{114.25}{GHz} -- or even beyond -- which is something that direct extension of 5G NR Rel-15 FR2 technology does not facilitate.
Most plausible technology enhancements include further increased subcarrier spacings to support wider channel bandwidths with fixed FFT size, and to enable the use of higher order modulations beyond \SI{100}{GHz} carrier frequencies. In addition, new single carrier waveforms to further reduce the PAPR with different modulation orders can be pursued to improve the power efficiency and coverage over the whole frequency range.

\bibliographystyle{IEEEtran}
\bibliography{IEEEabrv,references}

\end{document}